\documentclass[twocolumn,showpacs,amsmath,amssymb,prb,superscriptaddress]{revtex4}

\usepackage{graphicx}
\usepackage{bm}
\usepackage{epstopdf}
\begin{document}

\title{Supercurrent in superconducting graphene}
\author{ N.~B. Kopnin }
\affiliation{ Low Temperature Laboratory, Aalto University School
of Science and Technology,\\ PO Box 15100, FI-00076 AALTO,
Finland} \affiliation{ L.D.~Landau Institute for Theoretical
Physics, 117940 Moscow, Russia}
\author{E.~B. Sonin}
\affiliation{The Racah Institute of Physics, Hebrew University of
Jerusalem, Israel}

\date{\today}

\begin{abstract}
The problem of supercurrent in superconducting graphene is
revisited and the supercurrent is calculated within the mean-field
model employing the two-component wave functions on a honeycomb
lattice with pairing between different valleys in the Brillouin
zone. We show that the supercurrent within the linear
approximation in the order-parameter-phase gradient is always
finite even if the doping level is exactly zero.
\end{abstract}

\maketitle

\section{Introduction}

Recent exciting developments in transport experiments on graphene
\cite{Novoselov05} have stimulated theoretical and experimental
studies of possible superconductivity phenomena in this material.
Experimentally, there are both hints towards intrinsic
superconductivity \cite{Esquinazi08} and observations of
proximity-induced superconductivity in graphene layers
\cite{proxim}. Intrinsic superconductivity has been discussed
theoretically in the frameworks of phonon and plasmon mediated
mechanisms \cite{Uchoa07,GarcEsq09} whereas resonating valence
bond and density wave lattice models were proposed in Refs.
\onlinecite{Baskaran,Doniach07,CastroNeto05}. It was shown within
the BCS model\cite{CastroNeto05,Marino,KopninSonin08} that the
superconducting transition in the undoped graphene possesses a
quantum critical point at a finite interaction strength below
which the critical temperature vanishes. However, electrons in
graphene may become unstable towards formation of Cooper pairs for
any finite pairing interaction if doping shifts the Fermi level by
an amount $\mu$ away from the Dirac point
\cite{CastroNeto05,Uchoa07,Doniach07,KopninSonin08}. The effect of
fluctuations on the critical temperature of superconducting
transition in graphene has been studied in Ref.
\onlinecite{LoktevTurk09}. A number of unusual features of
superconducting state have been predicted, which are closely
related to the Dirac-like spectrum of normal state excitations. In
particular, the unconventional normal electron dispersion has been
shown to result in a nontrivial modification of Andreev reflection
\cite{been2} and Andreev bound states in Josephson junctions
\cite{been3} and vortex cores (see Ref. \onlinecite{KKMS09} and
references therein).

Nevertheless, there still remains a controversy regarding the most
fundamental property of superconducting graphene, i.e., the {\it
supercurrent}, no matter what the mechanism, intrinsic or
extrinsic, of the superconductivity is. In Ref.
\onlinecite{CastroNeto05} the supercurrent has been calculated
within the framework of the mean field model of superconducting
graphene \cite{CastroNeto05,been2} that assumes the Cooper pairing
between electrons belonging to the same sublattice in the
configurational space. According to Ref. \onlinecite{CastroNeto05}
the supercurrent calculated as a linear response to the phase
gradient of the order parameter disappears in undoped graphene
(i.e., zero shift of the chemical potential, $\mu=0$) at zero
temperature even if the order parameter $\Delta$ itself is finite.
However, a simpler model based on an effective Dirac type spectrum
of normal electrons \cite{KopninSonin08} demonstrates that the
supercurrent is always finite as long as superconductivity exists,
$\Delta \ne 0$. Though the surprising result\cite{CastroNeto05} of
``superconductivity without supercurrent'' is an alarming
indication by itself, the question may be raised, to which extent
this difference between the supercurrents is
model-dependent\cite{Castro-comment}, or, if not, what is then the
correct behavior of the supercurrent in the low doping limit, $\mu
\rightarrow 0$.

In the present paper we revisit this problem and calculate the
supercurrent again using the two-component mean field model of
superconductivity in graphene as formulated in Refs.
\onlinecite{CastroNeto05,been2}. We show that the supercurrent in
fact is {\it always finite}. Its value in the low doping limit
$\mu \ll |\Delta|$ is independent of whether the doping level is
exactly zero or not, in contrast to the claim of Ref.
\onlinecite{CastroNeto05}. This statement qualitatively agrees
with the conclusion drawn from the simple model suggested in Ref.
\onlinecite{KopninSonin08}.

The paper is organized as follows. In the next Section we outline
the model of superconductivity in graphene as formulated in Refs.
\onlinecite{CastroNeto05,been2} and introduce the basic quantities
relevant for further calculations. In Section
\ref{sec-current-linear} we calculate the supercurrent within the
linear approximation in the order-parameter phase gradient for
finite doping levels. The last Section
\ref{sec-current-degenerate} deals with the case of low doping
$\mu \ll |\Delta|$. Details of calculations are presented in
Appendix \ref{append1} and Appendix \ref{append2}.

\section{Bogoliubov--de Gennes--Dirac equations}

\begin{figure}[t]
\centering
\includegraphics[width=0.95\linewidth]{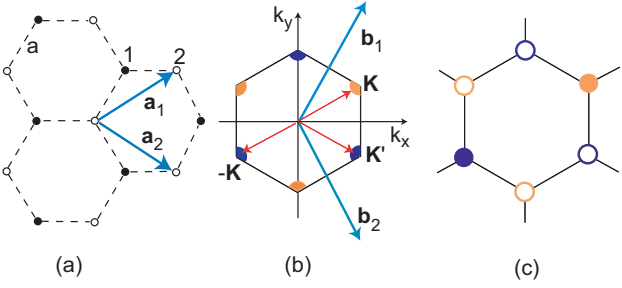}
\caption{(a) Unit cell with two sublattices 1 (black dots) and 2
(open dots), interatomic distance $a$, and the basis vectors ${\bf
a}_1$ and ${\bf a}_2$. (b) Brillouin zone with the reciprocal
lattice vectors ${\bf b}_1$ and ${\bf b}_2$. ${\bf K}$ and ${\bf
K}^\prime$ show the two non-equivalent Dirac corners (differently
shaded sectors) of the Brillouin zone; other corners are obtained
by shifting these two by integer linear combinations $n_1{\bf
b}_1+n_2{\bf b}_2$. (c) Dirac cone regions (circles) in the
extended zone scheme. Filled circles belong to the same zone,
while open circles are from other zones.}\label{fig-BZ}
\end{figure}

Transport properties of graphene associated with energies much
smaller than the band width are conveniently described by
equations of the Dirac type for two-component wave function whose
two components are envelopes of the true wave functions for two
sublattices in the configurational space, Fig. \ref{fig-BZ}(a),
near the so called Dirac points ${\bf K}$ or ${\bf K}^\prime$ in
the Brillouin zone of the reciprocal lattice, Fig. \ref{fig-BZ}(b)
(for more details see, for example, Refs.
\onlinecite{been2,CastroNeto08}).

A hole-like excitation $\Psi_{\bf K}^{(h)}$ in the valley
associated with the point ${\bf K}$ is the complex conjugated wave
function of a particle-like excitation in the valley $-{\bf K}$,
i.e., $ \Psi_{\bf K}^{(h)}= \Psi_{-{\bf K}}^* $. In what follows
we denote particle-like states by $ u$ while hole-like states by
$v$. The Bogoliubov--de Gennes equations have the form
\cite{been2,CastroNeto05}
\begin{eqnarray}
v_F {\bm \sigma}\cdot \left(-i{\bm \nabla}-\frac{e}{c}{\bf
A}\right) \hat u({\bf r})+\Delta \hat v({\bf r}) &=&(\epsilon+
\mu)
\hat u({\bf r})\ , \quad \label{DBdG-eK}\\
-v_F {\bm \sigma}\cdot \left(-i{\bm \nabla}+\frac{e}{c}{\bf
A}\right)  \hat v({\bf r})+\Delta^* \hat u({\bf r}) &=&(\epsilon
-\mu) \hat v({\bf r})\ . \quad \label{DBdG-hK}
\end{eqnarray}
The two-component wave functions are in a form of pseudo-spinors
\[
\hat u =\left( \begin{array}{c}u_1\\ u_2\end{array} \right)  , \;
\hat v =\left( \begin{array}{c}v_1\\ v_2\end{array} \right)  , \;
\hat u^\dagger = \left( u^*_1 , \; u^*_2\right)   , \; \hat
v^\dagger = \left( v^*_1, \; v^*_2\right) ,
\]
where the two components are the wave functions of electrons and
holes on two sublattices 1 and 2 in the honeycomb lattice, Fig.
\ref{fig-BZ}(a); $\hat {\bm \sigma}=(\hat \sigma _x\ , \;
\hat\sigma_y)$ are Pauli matrices in the pseudo-spin space:
\[
\hat \sigma _x= \left(\begin{array}{cc} 0 & 1 \\ 1 &
0\end{array}\right)\ , \; \hat \sigma _y= \left(\begin{array}{cc}
0 & -i \\ i & 0\end{array}\right)\ .
\]
Equations for the valley at the point ${\bf K}^\prime$ can be
obtained with the replacement $ u_1 \rightarrow u_2\ , \; v_1
\rightarrow v_2$.

Pairing of a particle $u$ in the valley ${\bf K}$ in the Brillouin
zone occurs with a hole $v$ at ${\bf K}$, i.e., with a particle in
the valley $-{\bf K}$. Since the points $-{\bf K}$ and ${\bf
K}^\prime$ are equivalent, ${\bf K}+{\bf K}^\prime={\bf b}_1+{\bf
b}_2$ where ${\bf b}_1$ and ${\bf b}_2$ are the vectors of the
reciprocal lattice, Fig. \ref{fig-BZ}(b), one may also say that
pairing is between particles from the valleys ${\bf K}$ and ${\bf
K}^\prime$. The model assumes that the order parameter is the same
for both sublattices,
\begin{equation}
\Delta =-V \sum_{{\bf p},\alpha} \left(1-2f_{{\bf
p},\alpha}\right) \hat v^\dagger _{{\bf p},\alpha} \hat u_{{\bf
p},\alpha}  \label{OP}
\end{equation}
where $\alpha$ labels four independent solutions of the
Bogoliubov--de Gennes equations with the momentum ${\bf p}$ (see
below) and $f_{{\bf p},\alpha}$ is the Fermi occupation number in
the state ${\bf p}, \alpha$. The sum runs over all states within
the Brillouin zone. We do not concentrate here on the specific
nature of the pairing interaction assuming that the pairing
potential may be either due to some intrinsic mechanism or due to
an interaction induced by a proximity to a usual superconductor.

The particle density is
\[
N=2 \sum_{{\bf p},\alpha}  \left[ f_{{\bf p},\alpha} \hat
u^\dagger _{{\bf p},\alpha}\hat u_{{\bf p},\alpha} +(1-f_{{\bf
p},\alpha})\hat v^\dagger _{{\bf p},\alpha}\hat v_{{\bf
p},\alpha}\right] \ .
\]
Factor 2 accounts for the true spin of electrons. The statistical
average of the current operator is
\begin{equation}
{\bf j} = 2 e v_F \sum_{{\bf p},\alpha}\left[ \hat u^\dagger
_{{\bf p},\alpha} \hat {\bm \sigma} \hat u_{{\bf p},\alpha}
f_{{\bf p},\alpha} - \hat v^\dagger _{{\bf p},\alpha} \hat {\bm
\sigma} \hat v_\alpha(1-f_{{\bf p},\alpha}) \right] \ .
\label{supercurrent-def}
\end{equation}
Sometimes the currents ${\bf j}_e $ and ${\bf j}_p $ are defined,
\begin{eqnarray}
{\bf j}_e&=&-e v_F\sum_{{\bf p},\alpha}\left[ \hat u^\dagger
_{{\bf p},\alpha} \hat {\bm \sigma} \hat u_{{\bf p},\alpha}+ \hat
v^\dagger _{{\bf p},\alpha} \hat {\bm \sigma} \hat
v_\alpha\right](1-2f_{{\bf p},\alpha}) \ , \quad \label{supercurrent}\\
{\bf j}_p &=&e v_F\sum_{{\bf p},\alpha} \left[ \hat
u^\dagger_\alpha \hat {\bm \sigma} \hat u_\alpha - \hat
v^\dagger_\alpha \hat {\bm \sigma} \hat v_\alpha\right]\ ,\quad
\end{eqnarray}
such that ${\bf j}={\bf j}_e+{\bf j}_p $. The current ${\bf j}_p$
is the quasiparticle flux, which vanishes in our spatially uniform
case (see below). The current ${\bf j}_e$ is the sum of currents
in each state, which may be not conserved separately in some
spatially inhomogeneous or non-equilibrium situations, but the
total current, however, is conserved ($ {\rm div}{\bf j}_e  =0 $)
taking into account the self-consistency equation \cite{tinkham}.

We will consider the case of zero magnetic field and look for the
solution in the form of  plane waves
\begin{equation}
\hat u_{\bf p} =\hat u e^{i({\bf p}+{\bf k}/2)\cdot {\bf r}}\ , \;
\hat v_{\bf p} =\hat v e^{i({\bf p}-{\bf k}/2)\cdot {\bf r}}\ ,
\label{hatu-p}
\end{equation}
assuming that  the order parameter $\Delta =|\Delta|e^{i{\bf
k}\cdot{\bf r}}$  corresponds to a moving condensate of Cooper
pairs. Equations (\ref{DBdG-eK}) and (\ref{DBdG-hK}) give
\begin{eqnarray}
v_F\hat {\bm \sigma}\cdot ({\bf p}+{\bf k}/2) \hat u +\Delta \hat
v =(E+\mu)\hat u \ ,
\label{dirBdG0-1} \\
-v_F\hat {\bm \sigma}\cdot ({\bf p}-{\bf k}/2) \hat v +\Delta^*
\hat u =(E-\mu)\hat v \ . \label{dirBdG0-2}
\end{eqnarray}

\subsection{Ground state}

Let us consider the ground state with zero current  (${\bf k}=0$).
Equations (\ref{dirBdG0-1}), (\ref{dirBdG0-2}) define four
linearly independent solutions. Let us introduce the spinors
\begin{equation}
\hat a_\uparrow = \frac{1}{\sqrt{2}}\left( \begin{array}{c}
\sqrt{\frac{p_x-ip_y}{p}} \\
\sqrt{\frac{p_x+ip_y}{p}}
\end{array}\right)\ , \;
\hat a_\downarrow = \frac{1}{\sqrt{2}} \left( \begin{array}{c}
\sqrt{\frac{p_x-ip_y}{p}} \\
- \sqrt{\frac{p_x+ip_y}{p}}
\end{array}\right), \label{spin-vects}
\end{equation}
which satisfy
\begin{equation}
(\hat {\bm \sigma}\cdot {\bf p} )\hat a_{\uparrow, \downarrow}
=\pm  p\, \hat a_{\uparrow, \downarrow} \ . \label{chiral1}
\end{equation}
The spinors $\hat a_\uparrow$ and $\hat a_\downarrow$ are
eigenstates of excitations in the normal graphene. We also
introduce vectors in the Nambu space,
\[
\check \psi = \left(\begin{array}{c} \hat u \\ \hat
v\end{array}\right)\ , \; \check \psi^+ =\left( \hat u^\dagger\ ,
\; \hat v^\dagger \right).
\]
Each component here is a pseudo-spinor. We find for the upper sign
in Eq. (\ref{chiral1})
\begin{equation}
E_{1,2}^{(0)}=\pm E_\uparrow \ , \; E_\uparrow =\sqrt{(v_Fp -
\mu)^2+|\Delta|^2} \ .  \label{E12}
\end{equation}
For ${\bf k}=0$ the order parameter is real $\Delta =|\Delta|$.
Therefore,
\begin{equation}
\left( \begin{array}{c} \hat u_1^{(0)} \\ \hat v_1^{(0)}
\end{array}\right) =
\left( \begin{array}{c} u_\uparrow \\  v_\uparrow
\end{array}\right) \hat a_\uparrow e^{i{\bf p}\cdot {\bf r}} ,
\left( \begin{array}{c} \hat u_2^{(0)} \\ \hat v_2^{(0)}
\end{array}\right) =
\left( \begin{array}{c} v_\uparrow \\ - u_\uparrow
\end{array}\right) \hat a_\uparrow e^{i{\bf p}\cdot {\bf r}}   . \label{u-v-1}
\end{equation}
For the lower sign in Eq. (\ref{chiral1}) we have
\begin{equation}
E_{3,4}^{(0)}=\pm E_\downarrow \ , \; E_\downarrow =\sqrt{(v_Fp +
\mu)^2+|\Delta|^2} \ , \label{E34}
\end{equation}
and
\begin{equation}
\left( \begin{array}{c} \hat u_3^{(0)} \\ \hat v_3^{(0)}
\end{array}\right) =
\left( \begin{array}{c} u_\downarrow \\
v_\downarrow
\end{array}\right) \hat a_\downarrow e^{i{\bf p}\cdot {\bf r}},
\left( \begin{array}{c} \hat u_4^{(0)} \\ \hat v_4^{(0)}
\end{array}\right) =
\left( \begin{array}{c} v_\downarrow \\ - u_\downarrow
\end{array}\right) \hat a_\downarrow e^{i{\bf p}\cdot {\bf r}}  . \label{u-v-2}
\end{equation}
Here
\begin{eqnarray}
u_\uparrow= \frac{1}{\sqrt{2}}\sqrt{
1+\frac{v_Fp-\mu}{E_\uparrow}}, \; v_\uparrow=
\frac{1}{\sqrt{2}}\sqrt{
1-\frac{v_Fp-\mu}{E_\uparrow}}\ ,\label{uv-up} \\
u_\downarrow= \frac{1}{\sqrt{2}}\sqrt{
1-\frac{v_Fp+\mu}{E_\downarrow}} , \; v_\downarrow=
\frac{1}{\sqrt{2}}\sqrt{ 1+\frac{v_Fp+\mu}{E_\downarrow}}\ .
\label{uv-down}
\end{eqnarray}
The different wave functions are orthogonal, $\check \psi^+_\alpha
\check \psi_\beta =\delta_{\alpha \beta}$. Equation (\ref{u-v-1})
goes over into Eq. (\ref{u-v-2}) under the transformation
$E\rightarrow -E$ and $\mu \rightarrow -\mu$. Using Eq.
(\ref{spin-vects}) one can check that
\begin{eqnarray}
\hat a^\dagger_\uparrow  \hat {\bm \sigma}  \hat a_\uparrow
&=&-\hat a^\dagger_\downarrow \hat {\bm \sigma} \hat a_\downarrow
={\bf p}/p \ ,
\label{diagon} \\
\hat a^\dagger_\uparrow  \hat {\bm \sigma}  \hat a_\downarrow
&=&-\hat a^\dagger_\downarrow \hat {\bm \sigma} \hat a_\uparrow
=i[{\bf z}_0\times{\bf p}]/p \ . \label{offdiagon}
\end{eqnarray}
where ${\bf z}_0$ is the unit vector in the $z$ direction
perpendicular to the graphene layer plane.

\section{Current-carrying state}
\label{sec-current-linear}

For the current-carrying state the solvability condition of  the
Bogoliubov--de Gennes equations (\ref{dirBdG0-1}),
(\ref{dirBdG0-2}) takes the form
\begin{eqnarray}
(E^2-\mu^2)^2-2|\Delta|^2(E^2-\mu^2)+|\Delta|^4
+2|\Delta|^2v_F^2{\bf p}_+{\bf p}_- \nonumber \\
-(E+\mu)^2v_F^2{\bf p}_-^2-(E-\mu)^2v_F^2{\bf p}_+^2 + v_F^4{\bf
p}_+^2 {\bf p}_-^2=0,~~ \label{cond-k}
\end{eqnarray}
where ${\bf p}_\pm ={\bf p}\pm {\bf k}/2$.

Equation (\ref{cond-k}) cannot be solved analytically for nonzero
${\bf k}$, except for the zero doping limit $\mu=0$. In the latter
case the solvability condition Eq.~(\ref{cond-k}) becomes
bi-quadratic and yields the energy spectrum
\begin{equation}
E_\pm^2=|\Delta|^2+v_F^2 (p^2+ k^2/4) \pm \sqrt{|\Delta|^2 v_F^2
k^2 +v_F^4 ({\bf p}\cdot {\bf k})^2}\ . \label{Emu0}
\end{equation}
In this limit, the Bogoliubov--de Gennes equations can also be
solved analytically (see Appendix \ref{undoped}). One sees that
the energy, Eq. (\ref{Emu0}), for $\mu =0$ does not have the usual
Doppler term proportional to the vector ${\bf k}$. This may lead
to a confusion \cite{Castro-comment} when calculating the
supercurrent.

\subsection{Linear response}

Let us consider the linear correction to the energy and to the
wave functions due to superconducting momentum ${\bf k}$ assuming
$v_F|{\bf k}|\ll \mu$. We put $E=E(0) +E^\prime$ where $E^\prime
\ll E(0)$ and $E(0)$ is the energy of one of the states with ${\bf
k}=0$ determined by Eqs. (\ref{E12}), (\ref{E34}). Within the
linear approximation in ${\bf k}$ we find from Eq. (\ref{cond-k})
for any finite $\mu \ne 0$
\[
E^\prime=\pm v_F ({\bf p}\cdot{\bf k})/2p \equiv \pm E_D
\]
for the upper (lower) sign in Eq. (\ref{chiral1}). Therefore,
corrections to the energies are
\begin{equation}
E_{1,2}^{(1)}=-E_{3,4}^{(1)}=E_D \ .\label{Ecorr1}
\end{equation}
The energy $E_D=(d\xi_{\bf p}/dp)({\bf k}/2) $ is the usual
Doppler shift for the normal-state energy $\xi_{\bf p}=v_Fp$.
Equation (\ref{Ecorr1}) coincides with the result of Ref.
\onlinecite{KopninSonin08} obtained in the linear approximation in
${\bf k}$. At the same time, it differs from the linear in ${\bf
k}$ term obtained from Eq. (\ref{Emu0}) for $\mu =0$. This means
that the undoped case $\mu =0$ requires a special consideration.
This will be done later in Section \ref{sec-current-degenerate}
(see also Appendix \ref{append2}).

First-order corrections to the wave functions can be found by
expanding the total functions in terms of the zero-order functions
$ \hat u_{\beta}^{(0)}$, $ \hat v_{\beta}^{(0)}$ given by Eqs.
(\ref{u-v-1}), (\ref{u-v-2}):
\begin{equation}
\check \psi _\alpha = \check \psi_\alpha ^{(0)}+\sum_{\beta\ne
\alpha} B_{\alpha \beta}\check\psi_\beta ^{(0)}\ .
\label{uv-total}
\end{equation}
Inserting this into Eqs. (\ref{DBdG-eK}), (\ref{DBdG-hK}) we find
\[
B_{\alpha\beta}=\frac{v_F\, \check \psi_{\beta}^{(0)+} (\hat {\bm
\sigma}\cdot {\bf k}) \check
\psi_{\alpha}^{(0)}}{2(E_{\alpha}^{(0)}-E_{\beta}^{(0)})} \ .
\]
One can check that $B_{\beta \alpha}=-B^*_{\alpha \beta}$. We find
$B_{12}=B_{21}=B_{34}=B_{43}=0$ while
\begin{eqnarray}
B_{13}&=&-B_{24}=- \frac{iv_F([{\bf p}\times {\bf k}]\cdot {\bf
z})}{2p} \frac{(u^*_\downarrow u_\uparrow +v^*_\downarrow
v_\uparrow)} {E_\uparrow -E_\downarrow} \ , \quad \label{C1324-lin}\\
B_{23}&=&B_{14}=\frac{iv_F([{\bf p}\times {\bf k}]\cdot {\bf
z})}{2p} \frac{(u^*_\downarrow v_\uparrow - v^*_\downarrow
u_\uparrow)} {E_\uparrow + E_\downarrow}\ .  \label{C2314-lin}
\end{eqnarray}
Therefore, the up-spin wave functions $\hat u^{(1,2)}$ contain
only corrections with the down-spin components $\hat u^{(3,4)}$,
and vice versa. Expansion Eq. (\ref{uv-total}) yields also the
corrections to the eigenenergies which coincide with Eq.
(\ref{Ecorr1}).

Using Eq. (\ref{uv-total}) one can show that the quasiparticle
current $ {\bf j}_{p}$ is zero. The supercurrent Eq.\
(\ref{supercurrent-def}) takes the form of Eq.
(\ref{supercurrent}), ${\bf j}={\bf j}_e$, which can be written as
\begin{equation}
{\bf j}=\int\frac{d^2p}{(2\pi)^2}\left[ {\bf j}_{{\bf K}}({\bf p})
+ {\bf j}_{-{\bf K}}({\bf p})\right] \ . \label{s-current}
\end{equation}
where
\begin{eqnarray}
{\bf j}_{\bf K}({\bf p}) =-e v_F  \sum_{\alpha =1}^4 \hat u_{{\bf
p},\alpha} ^{\dagger} \hat {\bm \sigma} \hat u_{{\bf p},\alpha}
\left[1-2f_{{\bf p},\alpha}\right]\ ,
\label{jK-gen}\\
{\bf j}_{-{\bf K}}({\bf p})=-e v_F  \sum_{\alpha =1}^4 \hat
v_{{\bf p},\alpha} ^{\dagger} \hat {\bm \sigma} \hat v_{{\bf
p},\alpha} \left[1-2f_{{\bf p},\alpha}\right]\ . \label{j-K-gen}
\end{eqnarray}
The term ${\bf j}_{\bf K}({\bf p})$ is the contribution from the
valley ${\bf K}$ in the Brillouin zone while ${\bf j}_{-{\bf
K}}({\bf p})$ is the contribution from valley $-{\bf K}$.
Therefore, Eq. (\ref{s-current}) in fact collects contributions
from the vicinity of the Dirac points at the opposite corners of
the entire Brillouin zone [shaded sectors in Fig.\ \ref{fig-BZ}(b)
or (c)].

Using Eq. (\ref{uv-total}) we obtain from Eq. (\ref{supercurrent})
in the linear approximation
\begin{widetext}
\begin{eqnarray}
{\bf j}&=&-e v_F \sum_{\alpha, {\bf p}}  \left[ \hat
u_{\alpha}^{(0)\dagger} \hat {\bm \sigma} \hat u_{\alpha}^{(0)} +
\hat v_{\alpha}^{(0)\dagger} \hat {\bm \sigma} \hat
v_{\alpha}^{(0)}
\right]\left[1-2f(E_{\alpha}^{(0)}+E_{\alpha}^{(1)})\right]\nonumber \\
&&-2e v_F {\rm Re} \sum_{\alpha\ne \beta , {\bf p}}
B_{\alpha\beta} \left[ \hat u_{\alpha}^{(0)\dagger} \hat {\bm
\sigma} \hat u_{\beta}^{(0)} + \hat v_{\alpha}^{(0)\dagger} \hat
{\bm \sigma} \hat v_{\beta}^{(0)}
\right]\left[1-2f(E_{\alpha}^{(0)})\right]\ . \label{lincurr1}
\end{eqnarray}
\end{widetext}

Equation (\ref{lincurr1}) contains the terms which diverge for
large $v_F p \gg |\Delta|, T$ because of  the contributions from
the Fermi sea of the states with negative energies, which extend
over the entire Brillouin zone including regions far from the
Dirac point. This divergence is spurious and can be eliminated
using two equivalent methods.

First, we note that the divergence of this kind is caused simply
by the fact that the overall shift of the particle momentum in the
Brillouin zone creates corrections to the wave functions which do
not decay as functions of the momentum far from the Dirac points.
Let us consider a change in the particle momentum ${\bf
p}\rightarrow {\bf p}+\delta {\bf p}$ everywhere in the Brillouin
zone. It will lead to the shift ${\bf p}\rightarrow {\bf p}+\delta
{\bf p}$ in the functions $u$ and, at the same time, to the shift
${\bf p}\rightarrow {\bf p}-\delta {\bf p}$ in $v$, because the
functions $v$ are associated with the complex conjugated wave
functions $u$ taken at the point $-{\bf K}$. In this way, the wave
functions used in Eq.\ (\ref{hatu-p}) contain corrections
associated with the overall shift ${\bf p}\rightarrow {\bf p}+{\bf
k}/2$ in the Brillouin zone. It is thus legitimate to
simultaneously change the momentum under the integral in Eq.\
(\ref{s-current}) or Eq.\ (\ref{lincurr1}) back to its original
value ${\bf p}$, i.e., to change the blind integration variable
${\bf p}\rightarrow {\bf p}- {\bf k}/2$ in the first and ${\bf
p}\rightarrow {\bf p}+ {\bf k}/2$ in second term. Excluding this
momentum shift we thus remove the diverging part, which is not
relevant to the supercurrent.

Within the linear approximation, it is sufficient to shift the
momenta in the zero-order term which comes from the first line of
Eq.\ (\ref{lincurr1}). The zero-order term yields
\begin{eqnarray*}
{\bf j}^{(0)}=\int\frac{d^2p}{(2\pi)^2}\left[ {\bf j}_{\bf
K}^{(0)}({\bf p}-{\bf k}/2)+ {\bf j}_{-{\bf K}}^{(0)}({\bf p}+{\bf
k}/2)\right]\nonumber \\
=\int\frac{d^2p}{(2\pi)^2}\left[ {\bf j}_{\bf K}^{(0)}({\bf p})+
{\bf j}_{-{\bf K}}^{(0)}({\bf p})- \left({\bf k}\cdot
\frac{\partial }{\partial {\bf p}}\right) {\bf j}_{\bf
K}^{(0)}({\bf p})\right]\ .
\end{eqnarray*}
Here ${\bf j}_{\bf K}^{(0)}({\bf p})$ and ${\bf j}_{-{\bf
K}}^{(0)}({\bf p})$ are the currents Eqs.\ (\ref{jK-gen}) and
(\ref{j-K-gen}) within the zero-order approximation in ${\bf k}$,
i.e., with the functions $\hat u_{\alpha}^{(0)}$ and $\hat
v_{\alpha}^{(0)}$ and the energies $E_\alpha ^{(0)}$ of states
Eqs.\ (\ref{E12})--(\ref{uv-down}) without a current. For these
states ${\bf j}_{\bf K}^{(0)}({\bf p})+{\bf j}_{-{\bf
K}}^{(0)}({\bf p})=0 $. As a result
\begin{equation}
{\bf j}^{(0)}= -\int \frac{ d\phi }{(2\pi)^2}  [({\bf p}\cdot {\bf
k}) {\bf j}_{\bf K}^{(0)}({\bf p})]_{p\gg \Delta, T}  \ .
\label{diverging}
\end{equation}
Here we transformed into the surface integral over a remote sphere
in the momentum space. At these momenta and energies, the current
in Eq.\ (\ref{diverging}) does not contain any information on the
superconducting properties of the material. This term compensates
the divergence of the corrections in the second line of that
equation.

Another way to remove the divergence in the second line of Eq.
(\ref{lincurr1}) would be to directly subtract from it the
normal-state current which is identically zero for the wave
functions specified by Eq. (\ref{hatu-p}). Indeed, as we already
mentioned, the diverging contributions to the current come from
the regions far from the Dirac point. Since, at these
quasiparticle momenta the energies greatly exceed the scales
relevant to the superconducting state, the corresponding
contributions to the current coincide with those in the normal
state. One can show that this regularization procedure leads to
the same result as Eq.\ (\ref{diverging}). The details of
calculations are given in Appendix \ref{append1}. The final
expression for the current becomes
\begin{widetext}
\begin{eqnarray}
{\bf j}&=&e v_F^2{\bf k} \int_0^\infty \frac{p\, dp}{2\pi}
\left[-\frac{1}{4T}\cosh^{-2}\frac{E_\uparrow
}{2T}-\frac{1}{4T}\cosh^{-2}\frac{E_\downarrow }{2T} +
\frac{\left| u_\uparrow ^*u_\downarrow
+v_\uparrow^*v_\downarrow\right|^2}{E_\uparrow
-E_\downarrow}\left( \tanh\frac{E_\uparrow}{2T}
-\tanh\frac{E_\downarrow}{2T} \right)\right.\nonumber \\
&&\left.+\frac{1}{v_F p}- \frac{\left| v_\uparrow ^*u_\downarrow
-u_\uparrow ^*v_\downarrow\right|^2}{E_\uparrow
+E_\downarrow}\left( \tanh\frac{E_\uparrow}{2T}
+\tanh\frac{E_\downarrow}{2T} \right) \right] \ .
\label{s-current-fin}
\end{eqnarray}
\end{widetext}
It is worthwhile to note that the problem of spurious divergent
terms in the expression for the supercurrent is rather general. In
particular, it was discussed (and resolved similarly) for the
superfluid excitonic current in graphene bilayers\cite{MacD}.

We evaluate Eq. (\ref{s-current-fin}) for low temperatures, $T\ll
|\Delta|$. Since $ E_\uparrow , E_\downarrow >|\Delta|$, the first
line in Eq. (\ref{s-current-fin}) vanishes at $T=0$. The
supercurrent becomes
\begin{equation}
{\bf j}=\frac{ e{\bf k}}{2\pi}\left[ \sqrt{\mu ^2+|\Delta|^2} +
\frac{|\Delta|^2}{|\mu|} \ln \left( \frac{|\mu| +\sqrt{\mu
^2+|\Delta|^2}}{|\Delta|}\right)\right] . \label{currentT0}
\end{equation}

For $\mu \gg |\Delta|$ we have
\begin{equation}
{\bf j}= e|\mu|{\bf k}/2\pi \ . \label{lin-curr-bigmu}
\end{equation}
For $\mu \ll |\Delta|$ we find
\begin{equation}
{\bf j}=e|\Delta|{\bf k}/\pi\ .\label{lin-curr-mu0}
\end{equation}
This result formally holds within the linear approximation which
assumes $v_F k \ll \mu$. Therefore, in Eq. (\ref{lin-curr-mu0})
one would have to put $k\rightarrow 0$ first and then assume $\mu
\ll |\Delta|$. In the next section we demonstrate that Eq.
(\ref{lin-curr-mu0}) is in fact always valid provided $|\mu| \ll
|\Delta|$ and $v_F k \ll |\Delta|$ irrespectively of the relation
between $v_F k$ and $\mu$.

\subsection{Low doping limit} \label{sec-current-degenerate}

Consider now the limit of small $\mu$ when the zero order state is
degenerate because $E^{(0)}_1=E^{(0)}_3 =E_0$ and
$E^{(0)}_2=E^{(0)}_4=-E_0 $ where
\begin{equation}
E_0=\sqrt{(v_F p)^2+|\Delta|^2} \ . \label{E02-E04}
\end{equation}
In what follows we demonstrate that despite the absence of the
Doppler term in its usual form, there still is a finite linear in
${\bf k}$ supercurrent down to zero temperature (contrary to the
result of Ref. \onlinecite{CastroNeto05}).

The true wave functions satisfy
\[
\check H \check \psi_\alpha =E_\alpha \check\psi_\alpha
\]
where $ \check H=\check H^{(0)} +\check H^{(1)} $ and
\begin{eqnarray*}
\check H^{(0)}&=& \left(\begin{array}{cc} v_F \hat{\bm
\sigma}\cdot {\bf p} & |\Delta|\\ |\Delta| &  -v_F \hat{\bm
\sigma}\cdot {\bf
p}\end{array}\right)\ , \\
\check H^{(1)}&=& \left(\begin{array}{cc} \frac{1}{2}v_F \hat{\bm
\sigma}\cdot {\bf k} -\mu & 0\\0 & \frac{1}{2}v_F \hat{\bm
\sigma}\cdot {\bf k} +\mu\end{array}\right)\ .
\end{eqnarray*}
We assume $v_F k, \mu \ll E_0$.

Let us expand the true wave function into the zero order
orthonormal wave functions $\check\psi_\alpha ^{(0)}$,
\begin{equation}
\check \psi_\alpha =\sum_\beta
C_{\alpha\beta}\check\psi_\beta^{(0)}\ , \label{expansion1}
\end{equation}
satisfying the zero-order equation
\[
\check H^{(0)} \check\psi_\alpha ^{(0)} =E_\alpha^{(0)}
\check\psi_\alpha^{(0)} \ .
\]
The zero-order wave functions have now the form of Eqs.
(\ref{u-v-1}), (\ref{u-v-2}) with $u_\uparrow=v_\downarrow\equiv
u$ and $u_\downarrow=v_\uparrow\equiv v $ where
\begin{equation}
u=\frac{1}{\sqrt{2}}\left[ 1+ \frac{v_Fp}{E_0}\right]^{1/2} \ ,\;
v=\frac{1}{\sqrt{2}}\left[ 1- \frac{v_Fp}{E_0}\right]^{1/2}\ .
\label{uv0}
\end{equation}
The expansion coefficients satisfy
\begin{equation}
C_{\alpha \gamma}\left[ E_\alpha -E_\gamma ^{(0)}\right]=
\sum_\beta C_{\alpha \beta} H_{\gamma \beta} \label{Cexpans1}
\end{equation}
where $ H_{\gamma \beta}\equiv \left< \check \psi_\gamma ^{(0)+}
\check H^{(1)} \check \psi_\beta ^{(0)}\right> $.


Consider the state $\alpha =1$. Since the difference $
E_1-E_1^{(0)}=E_1-E_3^{(0)}\equiv \delta E_1 $ in Eq.
(\ref{Cexpans1}) is small, the coefficients $C_{12}$ an $C_{14}$
are proportional to the perturbation, while $C_{11}$ and $C_{13}$
are of the order unity. The coefficients $C_{11}^{(0)} $,
$C_{13}^{(0)} $ in the leading approximation satisfy the secular
equations (\ref{secul11}), (\ref{secul13}) (see Appendix
\ref{append2}) which yield $\delta E_{1,3} =\mp \tilde E_1$,
\begin{equation}
\tilde E_1= \sqrt{\left(\frac{\mu v_Fp}{E_0}-E_D\right)^2+
\frac{v_F^2 [{\bf p}\times {\bf
k}]^2}{4p^2}\frac{|\Delta|^2}{E_0^2}} \label{deltaE13}
\end{equation}
and
\begin{eqnarray}
C_{11}^{(0)}&=&C_{33}^{(0)}=\frac{1}{\sqrt{2}}\sqrt{1+\frac{\frac{\mu
v_F p}{E_0}-E_D} {\tilde E_1 }}\ ,
\label{C1133}\\
C_{13}^{(0)}&=&C_{31}^{(0)}=\frac{i{\rm sign} ([{\bf p}\times {\bf
k}]{\bf z})}{\sqrt{2}}\sqrt{1-\frac{\frac{\mu v_F p}{E_0}-E_D}
{\tilde E_1 }} \ . \qquad \label{C1331}
\end{eqnarray}
The coefficients obey the normalization
$|C_{11}^{(0)}|^2+|C_{13}^{(0)}|^2=
|C_{33}^{(0)}|^2+|C_{31}^{(0)}|^2=1$. In the same way we find
$\delta E_{2,4} =\pm \tilde E_2 $. The energy correction $\tilde
E_2$ and the coefficients $C_{22}^{(0)}=C_{44}^{(0)}$ and
$C_{24}^{(0)}=C_{42}^{(0)} $ are obtained from $\tilde E_1$,
$C_{11}^{(0)}$, and $C_{13}^{(0)} $, respectively, by replacing
${\bf p} \rightarrow -{\bf p}$.

For $k/p \ll \mu /|\Delta|$ we have
\[
\tilde E_{1,2}= \frac{\mu v_Fp}{E_0}\mp E_D
\]
and $ C_{11}=C_{33}= C_{22}=C_{44}=1 \ , \;
C_{13}=C_{31}=C_{24}=C_{42}=0 $ which agrees with the result of
the linear approximation in $k$. The coefficients $C_{ik}$ taken
for $v_Fk \ll \mu$ coincide with Eqs.
(\ref{C1324-lin})-(\ref{C2314-lin}) in the limit $\mu \ll E$. In
the limit $\mu =0$ we have
\[
\tilde E_1=\tilde E_2 \equiv \tilde E =(v_F p/E_0)\sqrt{E_D^2
+k^2|\Delta|^2/4p^2}
\]
which agrees with Eq. (\ref{Emu0}).

Now we insert all four solutions into the expression  Eq.
(\ref{supercurrent}) for the current. Removing the divergence by
subtracting the current in the normal state we find
\begin{eqnarray}
{\bf j}= ev_F^2{\bf k} \int_0^\infty \frac{p\,
dp}{2\pi}\left[-\left(1 +\frac{|\Delta|^2}{E_0^2}
\right)\frac{d}{dE_0} \tanh
\frac{E_0}{2T}\right. && \nonumber \\
\left. +\frac{1}{v_Fp} -\frac{v_F^2 p^2}{E_0^3}  \tanh
\frac{E_0}{2T}\right] &.& \quad  \label{current-deg-T}
\label{currDeg}
\end{eqnarray}
It does not depend on $\mu$.

Consider low temperatures, $T\ll |\Delta|$. The first term
vanishes and we obtain
\begin{equation}
{\bf j} =e v_F^2 {\bf k} \int_0^\infty \frac{p\, dp}{2\pi} \left[
\frac{1}{v_F p} -\frac{v_F^2 p^2}{E_0^3}\right]=\frac{
e|\Delta|{\bf k}}{\pi} \label{curr-mu0}
\end{equation}
which is the same result as in the linear approximation, Eq.
(\ref{lin-curr-mu0}).

\section{Discussion and comparison}

As we already mentioned in the Introduction,  the present paper
studies the model \cite{CastroNeto05,been2} that assumes Cooper
pairing between electrons (holes) belonging to the same sublattice
in the configurational space. This is evident, in fact, from the
self-consistency equation Eq.~(\ref{OP}), which contains the
scalar product of the spinors $\hat u$ and $\hat v$. However,
other scenarios of the superconducting pairing are possible, as
well. In particular, one can use the approach which is based on
the Landau Fermi-liquid theory which operates with the
quasiparticles corresponding to the eigenstates of the
normal-state Hamiltonian [spinors  $\hat a_\uparrow$ and $\hat
a_\downarrow$ in Eq.~(\ref{chiral1})]. All essential properties
are then derived based on the quasiparticle energy spectrum. The
Fermi-liquid approach (with some variations) was used in
Refs.~\onlinecite{Baskaran,Doniach07,Marino,KopninSonin08,LoktevTurk09}.
The dilemma of ``intra-sublattice  interaction only''  vs.
``interaction between true quasiparticles of the normal graphene''
was also discussed in connection with other collective modes in
graphene\cite{baskaran2}. In general, this dilemma can be resolved
only on the basis of the detailed microscopic analysis of the
particular interaction mechanism.

Nevertheless, the main outcome of our analysis is that the
superconducting behaviors calculated within the two aforementioned
approaches are qualitatively very similar, though intra-sublattice
interaction requires a more involved algebra (4-fold matrices
rather than 2-fold matrices in the Fermi-liquid approach).
Quantitatively, however, Eq. (\ref{currentT0}) is slightly
different from the result of Ref. \onlinecite{KopninSonin08}. In
particular, the current in the limit $\mu \gg |\Delta|$, Eq.\
(\ref{lin-curr-bigmu}), is twice as large as in Ref.
\onlinecite{KopninSonin08}. This factor 2 appears simply because
the model with two times smaller number of the degrees of freedom
(only one valley with a Dirac cone in the Brillouin zone) has been
considered in the cited paper. For large $\mu$, there should
otherwise be no difference in the superconducting properties,
since the both models practically coincide with that for usual
superconductors. However, the low-$\mu$ limit differs already by
factor 4. The latter is obviously a manifestation of a more subtle
difference between these two models.

Apart from this numerical difference, the global features of the
superconducting graphene are insensitive to the choice of the
pairing model. The main message of Ref. \onlinecite{KopninSonin08}
is confirmed that the supercurrent and the superconducting
electron density are finite at any doping level for all
temperatures below the critical temperature; in particular, they
do not disappear in the limit $\mu =0$ contrary to the claim of
Ref. \onlinecite{CastroNeto05}. We have in fact shown that the
low-doping limit $\mu \rightarrow 0$, being degenerate in the
excitation energies, is not any special in the sense of the
supercurrent: The supercurrent obtained within the linear
approximation in the gradient of the order parameter phase, $k \ll
|\Delta|/v_F$, is the same irrespectively of the relation between
$v_Fk$ and $\mu$. The crucial difference between the
superconducting graphene and the usual BCS superconductor is that
the supercurrent density in the low-doping limit at $T=0$ is
proportional to the order parameter $\Delta$ rather than to the
total electron density.

\acknowledgments

This work was supported by the grant of the Israel Academy of
Sciences and Humanities, by the Russian Foundation for Basic
Research under grant 09-02-00573-a, by the Program ``Quantum
Macrophysics'' of the Russian Academy of Sciences, and by the
Academy of Finland Centers of Excellence Program. NBK thanks for
hospitality the Racah Institute of Physics of the Hebrew
University of Jerusalem, where a part of this work has been
performed.

\appendix

\section{Current in the linear approximation}\label{append1}

We start with Eq. (\ref{lincurr1}). Using Eqs. (\ref{u-v-1}),
(\ref{u-v-2}), and (\ref{C1324-lin}), (\ref{C2314-lin}) we find
after averaging over momentum directions in the second line
\begin{widetext}
\begin{eqnarray}
{\bf j}&=&-e v_F \sum_{{\bf p}} \frac{{\bf p}}{p}
\left[\tanh\frac{E_\uparrow +E_D}{2T} - \tanh\frac{E_\uparrow -
E_D}{2T} + \tanh\frac{E_\downarrow +
E_D}{2T}-\tanh\frac{E_\downarrow -E_D}{2T} \right]
\nonumber \\
&&-  ev_F^2 {\bf k} \sum_{{\bf p}} \left[ \frac{\left| u_\uparrow
^*u_\downarrow +v_\uparrow^*v_\downarrow\right|^2}{E_\uparrow
-E_\downarrow}\left( \tanh\frac{E_\uparrow}{2T}
-\tanh\frac{E_\downarrow}{2T} \right)+ \frac{\left| v_\uparrow
^*u_\downarrow -u_\uparrow ^*v_\downarrow\right|^2}{E_\uparrow
+E_\downarrow}\left( \tanh\frac{E_\uparrow}{2T}
+\tanh\frac{E_\downarrow}{2T} \right) \right]\ .
\label{current-linexp}
\end{eqnarray}
\end{widetext}

This expression formally diverges for large $v_F p$ due to
contribution from regions far from the Dirac points. As we
discussed already in Section \ref{sec-current-linear} we remove
this spurious divergence by transforming the zero-order terms.
With Eqs. (\ref{uv-up}), (\ref{uv-down}) we have in the zero-order
approximation
\begin{eqnarray}
\!\!\!&&\!\!\!{\bf j}_{{\bf K}}^{(0)}({\bf p})=
-{\bf j}_{-{\bf K}}^{(0)}({\bf p})\nonumber \\
\!\!\!&&\!\!\!=-e v_F \left[ \frac{v_Fp-\mu}{E_\uparrow}
\tanh\frac{E_\uparrow}{2T}+ \frac{v_Fp+\mu}{E_\downarrow}
\tanh\frac{E_\downarrow}{2T}\right] \frac{\bf p}{p} \ . \qquad
\label{jK}
\end{eqnarray}
The contribution from the zero-order term has the form of Eq.
(\ref{diverging}) of surface integral over a remote sphere in the
momentum space. Using Eq. (\ref{jK}) we find
\begin{eqnarray}
{\bf j}^{(0)}= -\int \frac{ d\phi }{(2\pi)^2}  [({\bf p}\cdot {\bf
k})
{\bf j}_{\bf K}^{(0)}({\bf p})]_{p\rightarrow \infty} \nonumber \\
= ev_F^2 {\bf k} \int_0^\infty \frac{p\, dp}{2\pi}\left[
\frac{1}{v_Fp}\right] \ . \label{diverging1}
\end{eqnarray}
When added to  Eq. (\ref{current-linexp}), this compensates the
diverging terms there. As a result, we obtain the converging
expression, Eq. (\ref{s-current-fin}). The same result can be
obtained if we subtract the normal current, i.e., Eq.
(\ref{current-linexp}) for $\Delta =0$.

For $T\ll |\Delta|$ the terms $\cosh^{-2}(E_{\uparrow,
\downarrow}/2T)$ are small while $\tanh
(E_{\uparrow,\downarrow}/2T)=1$. Therefore, the first line in Eq.
(\ref{s-current-fin}) vanishes. The current becomes
\begin{equation}
{\bf j}= ev_F^2 {\bf k} \int_0^\infty \frac{p\,
dp}{2\pi}\left[\frac{1}{v_F p}- \frac{2\left| v_\uparrow
^*u_\downarrow -u_\uparrow ^*v_\downarrow\right|^2}{E_\uparrow
+E_\downarrow} \right]\ . \label{current1}
\end{equation}
Calculating the current with help of Eqs. (\ref{uv-up}),
(\ref{uv-down}) we obtain Eq. (\ref{currentT0}).

\section{Current in the degenerate case}\label{append2}

\subsection{Wave functions for weak doping $\mu \ll \Delta$}

Consider Eq. (\ref{Cexpans1}) for the state $\alpha =1$. We have
within the linear approximation in $H^{(1)}$
\begin{eqnarray}
C_{11}\left[ E_1 -E_1 ^{(0)}\right]=\sum_\beta C_{1
\beta}H_{1\beta} \ , \label{lin11}\\
C_{13}\left[ E_1 -E_3 ^{(0)}\right]=\sum_\beta C_{1
\beta}H_{3\beta}\ , \label{lin13}
\end{eqnarray}
and
\begin{eqnarray}
C_{12}\left[ E_1 -E_2 ^{(0)}\right]=\sum_\beta C_{1
\beta}H_{2\beta}\ , \label{lin12}\\
C_{14}\left[ E_1 -E_4 ^{(0)}\right]=\sum_\beta C_{1
\beta}H_{4\beta}\ . \label{lin14}
\end{eqnarray}
Since
\[
E_1-E_1^{(0)}=E_1-E_3^{(0)}\equiv \delta E_1
\]
are small, in Eqs. (\ref{lin11}) and (\ref{lin13}) we can take the
coefficients in the zero order approximation. As a result $C_{12}$
an $C_{14}$ are small (i.e., are proportional to the
perturbation), while $C_{11}$ and $C_{13}$ are of the order unity.
We put $C_{11}=C_{11}^{(0)} $, $C_{13}=C_{13}^{(0)} $ while
$C_{12}=C_{12}^{(1)} $, $C_{14}=C_{14}^{(1)} $, and find up to the
first order terms
\begin{eqnarray}
C_{11}^{(0)}\delta E_1^{(1)} =C_{11}^{(0)}H_{11} +
C_{13}^{(0)}H_{13} \ , \label{secul11}\\
C_{13}^{(0)}\delta E_1^{(1)} =C_{11}^{(0)}H_{31}+
C_{13}^{(0)}H_{33} \ , \label{secul13}
\end{eqnarray}
while
\begin{eqnarray}
2 E_0 C_{12}^{(1)}=C_{11}^{(0)}H_{21} + C_{13}^{(0)}H_{23}\ ,  \label{corr11} \\
2 E_0 C_{14}^{(1)}=C_{11}^{(0)}H_{41} + C_{13}^{(0)}H_{43}\ .
\label{corr12}
\end{eqnarray}
The similar equations are obtained for the other state $\alpha =3$
which belongs to the same energy $E_0$.

We have
\begin{eqnarray*}
H_{\alpha \beta}= \frac{v_F }{2}\left(\hat u^{+(0)}_\alpha
\hat{\bm \sigma}\cdot {\bf k}\hat u_\beta^{(0)} + \hat
v^{+(0)}_\alpha
\hat{\bm \sigma}\cdot {\bf k}\hat v_\beta^{(0)}\right) \\
-\mu\left( \hat u^{+(0)}_\alpha \hat u_\beta^{(0)} - \hat
v^{+(0)}_\alpha \hat v_\beta^{(0)}\right)\ .
\end{eqnarray*}
Therefore
\begin{eqnarray*}
H_{11}&=&-H_{33}=\frac{v_F {\bf p}\cdot {\bf k}}{2p} - \frac{\mu
v_Fp}{E_0}\ , \\
H_{13}&=&-H_{31}=\frac{iv_F ([{\bf z}_0\times {\bf p}]{\bf
k})}{2p}\frac{|\Delta|}{E_0}\ ,
\end{eqnarray*}
and
\begin{eqnarray}
H_{21}&=&H_{21}=H_{43}=H_{34}=-\mu\frac{|\Delta|}{E_0}\ , \label{H21}\\
H_{23}&=&-H_{32}=-i\frac{v_Fp}{E_0} \frac{v_F([{\bf
p}\times {\bf k}]{\bf z})}{2p}\ ,  \label{H23}\\
H_{41}&=&-H_{14}=-i\frac{v_Fp}{E_0} \frac{v_F([{\bf p}\times {\bf
k}]{\bf z})}{2p}\ . \label{H41}
\end{eqnarray}

Secular equations (\ref{secul11}), (\ref{secul13}) determine
$\delta E_{1,3}$ together with the coefficients $C^{(0)}_{11}$,
$C^{(0)}_{13}$. The first-order corrections to $C_{11}$ and
$C_{13}$ are found from Eqs. (\ref{lin11}) and (\ref{lin13})
written up to the second-order terms. Using Eqs. (\ref{H21}),
(\ref{H23}), and (\ref{H41}) we obtain $ C_{11}^{(1)}=
C_{13}^{(1)}=0 $.  The coefficients $C^{(1)}_{12}$ and
$C^{(1)}_{14}$ are determined by Eqs. (\ref{corr11}) and
(\ref{corr12}). In the same way we find $ C_{31}^{(1)}=
C_{33}^{(1)}=0 $ together with the coefficients $C^{(1)}_{32}$ and
$C^{(1)}_{34}$. Calculations for the states $\alpha =2,4$ can be
done in exactly the same way.

Using the obtained coefficients we can rewrite Eq.
(\ref{supercurrent}) for the current in the form
\begin{eqnarray}
{\bf j}&=&-ev_F \sum _{\bf p}\left[ A_1^{(0)} \left(\tanh
\frac{E_0-\tilde E_1}{2T} - \tanh \frac{E_0+\tilde E_1}{2T}\right)
\right. \nonumber \\
&&- A_2^{(0)} \left(\tanh \frac{E_0-\tilde E_2}{2T} -
\tanh \frac{E_0+\tilde E_2}{2T}\right) \nonumber \\
&& +\left. 2\left(A_1^{(1)}-A_2^{(1)}\right)\tanh
\frac{E_0}{2T}\right] \ , \label{deg-curr1}
\end{eqnarray}
where
\begin{eqnarray*}
A_1^{(0)}&=&\left( |C_{11}^{(0)}|^2 -|C_{13}^{(0)}|^2\right) \hat
a^\dagger_\uparrow \hat {\bm \sigma}\hat a_\uparrow \nonumber \\
&&+2uv \left( C^{(0)*}_{11}C^{(0)}_{13}
-C^{(0)*}_{13}C_{11}^{(0)}\right)
\hat a^\dagger_\uparrow \hat {\bm \sigma}\hat a_\downarrow \ ,\quad  \label{A10} \\
A_2^{(0)}&=&\left( |C_{22}^{(0)}|^2 -|C_{24}^{(0)}|^2\right) \hat
a^\dagger_\uparrow \hat {\bm \sigma}\hat a_\uparrow \nonumber \\
&&+2uv \left( C^{(0)*}_{22}C_{24}^{(0)}
-C_{24}^{(0)*}C_{22}^{(0)}\right) \hat a^\dagger_\uparrow \hat
{\bm \sigma}\hat a_\downarrow \ , \quad \label{A20}
\end{eqnarray*}
and
\begin{eqnarray*}
A_1^{(1)}&=&(u^2-v^2)E_0^{-1} H_{23}\hat
a^\dagger_\uparrow \hat {\bm \sigma}\hat a_\downarrow \ , \label{A11}\\
A_2^{(1)}&=&(u^2-v^2)E_0^{-1} H_{14}\hat a^\dagger_\uparrow \hat
{\bm \sigma}\hat a_\downarrow \ . \label{A21}
\end{eqnarray*}


The current in Eq. (\ref{deg-curr1}) takes the form
\begin{eqnarray}
{\bf j}&=&2e \sum _{\bf p}\left[\frac{\partial \tilde
E_1}{\partial {\bf k}}  \left(\tanh \frac{E_0-\tilde E_1}{2T} -
\tanh \frac{E_0+\tilde E_1}{2T}\right) \right. \nonumber \\
&&+\left. \frac{\partial \tilde E_2}{\partial {\bf k}} \left(\tanh
\frac{E_0-\tilde E_2}{2T} -
\tanh \frac{E_0+\tilde E_2}{2T}\right)\right.\nonumber \\
&&-\left. \frac{v_F^2 p^2}{E_0^3} \frac{v_F^2[{\bf p}\times {\bf
k}]\times {\bf p}}{p^2}\tanh \frac{E_0}{2T}\right] .
\label{current-degen1}
\end{eqnarray}
For $v_Fk \ll \mu$ the this equation coincides with Eq.
(\ref{current-linexp}) taken for $\mu \ll \Delta$. Expanding it in
small $\tilde E_1$ and $\tilde E_2$ we find
\begin{eqnarray*}
{\bf j}&=&-2e \sum _{\bf p}\left[\frac{\partial }{\partial {\bf
k}}\left( \tilde E_1^2+\tilde E_2^2\right)\frac{d}{dE_0} \tanh
\frac{E_0}{2T}\right. \\
&&+\left. \frac{v_F^2 p^2}{E_0^3} \frac{v_F^2[{\bf p}\times {\bf
k}]\times {\bf p}}{p^2}\tanh \frac{E_0}{2T}\right]
\end{eqnarray*}
which yields after integration over the momentum angle
\begin{eqnarray}
{\bf j}=- ev_F^2{\bf k} \sum_{\bf p}\left[\left(1
+\frac{|\Delta|^2}{E_0^2} \right)\frac{d}{dE_0} \tanh
\frac{E_0}{2T}\right. && \nonumber \\
\left. +\frac{v_F^2 p^2}{E_0^3}  \tanh \frac{E_0}{2T}\right]&.&
\quad \label{curent-fin-allmu}
\end{eqnarray}
Subtraction of the normal-state current returns us to Eq.
(\ref{currDeg}).

\subsection{Undoped graphene}  \label{undoped}

The Bogoliubov--de Gennes equations can be solved exactly for the
undoped case $\mu =0$ where the dispersion equation (\ref{cond-k})
becomes bi-quadratic with the energy spectrum Eq.~(\ref{Emu0}). We
shall write down the solution  of the Bogoliubov--de Gennes
equations for the vector ${\bf k}$ parallel to the axis $ x$
($k=k_x$). Then
\begin{equation}
E_\pm^2=(\sqrt{|\Delta|^2+v_F^2p_x^2}\pm v_Fk/2)^2+v_F^2p_y^2,
\end{equation}
and one may check by substitution that the four ortho-normalized
solutions of the Bogoliubov--de Gennes equations are given by the
spinors
\begin{eqnarray}
\hat u  &=& { 1 \over 2 } \sqrt{ 1\pm {p_x\over
\sqrt{|\Delta|^2/v_F^2+p_x^2}} }\left( \begin{array}{c}
\pm  {E_\pm\over |E_\pm|}e^{-i\phi_\pm /2} \\
e^{i\phi_\pm /2}
\end{array}\right) , \nonumber  \\
\hat v  &=& { 1 \over 2 } \sqrt{ 1\mp {p_x\over
\sqrt{|\Delta|^2/v_F^2+p_x^2}} }\left( \begin{array}{c}
\pm  e^{-i\phi_\pm /2} \\
{E_\pm\over |E_\pm|}e^{i\phi_\pm /2}
\end{array}\right),
\label{sol}\end{eqnarray} where the phase factors are given by
\[
e^{i\phi_\pm }=\frac{\sqrt{|\Delta|^2/v_F^2+p_x^2}\pm k/2\pm
ip_y}{\sqrt{(\sqrt{|\Delta|^2/v_F^2+p_x^2}\pm k/2)^2+p_y^2}}.
\]

Using Eq. (\ref{sol}) one finds the terms which determine the $x$
component of the supercurrent:
\begin{widetext}
\begin{eqnarray*}
\hat u^\dagger  \hat  \sigma_x   \hat u = \pm{ 1 \over 2 }
{E_\pm\over |E_\pm|}\sqrt{ 1\pm {p_x\over
\sqrt{|\Delta|^2/v_F^2+p_x^2}} }\cos \phi_\pm
\approx \pm { 1 \over 2 } {E_\pm\over |E_\pm|}\sqrt{ 1\pm
{p_x\over \sqrt{|\Delta|^2/v_F^2+p_x^2}} }\left(1\pm {kp_y^2\over
(|\Delta|^2/v_F^2+p^2)^{3/2}}  \right) ,\\
\hat v^\dagger  \hat \sigma_x  \hat v = \pm{ 1 \over 2 }
{E_\pm\over |E_\pm|}\sqrt{ 1\pm {p_x\over
\sqrt{|\Delta|^2/v_F^2+p_x^2}} }\cos \phi_\pm
\approx \pm { 1 \over 2 } {E_\pm\over |E_\pm|}\sqrt{ 1\mp
{p_x\over \sqrt{|\Delta|^2/v_F^2+p_x^2}} }\left(1\pm {kp_y^2\over
(|\Delta|^2/v_F^2+p^2)^{3/2}}  \right),
\end{eqnarray*}
\end{widetext}
where finally we keep only terms linear in $k$. Collecting all the
terms in the expression Eq. (\ref{supercurrent}) we arrive at
Eq.~(\ref{curent-fin-allmu}).

\end{document}